\documentclass[twocolumn,tighten]{aastex63}
\usepackage{mathtools}
\usepackage{txfonts} 
\usepackage{hyperref}
\usepackage{xcolor}
\usepackage{amsmath}

\defcitealias{PaperI}{EHTC~I}
\defcitealias{PaperII}{EHTC~II}
\defcitealias{PaperIII}{EHTC~III}
\defcitealias{PaperIV}{EHTC~IV}
\defcitealias{PaperV}{EHTC~V}
\defcitealias{PaperVI}{EHTC~VI}
\defcitealias{PaperVII}{EHTC~VII}
\defcitealias{PaperVIII}{EHTC~VIII}
\defcitealias{Narayan_2021}{N21}
\defcitealias{Gelles_2021}{G21}
\defcitealias{Palumbo_2022}{P22}

\newcommand{\m}{M87*}
\newcommand{\s}{Sgr\,A*}

\begin{document}                                              

\title{Demonstrating Photon Ring Existence with Single-Baseline Polarimetry}
\shorttitle{Single-Baseline Photon Ring Polarimetry}

\author[0000-0002-7179-3816]{Daniel~C.~M.~Palumbo}
\email{daniel.palumbo@cfa.harvard.edu}
\affiliation{Center for Astrophysics $\vert$ Harvard \& Smithsonian, 60 Garden Street, Cambridge, MA 02138, USA}
\affiliation{Black Hole Initiative at Harvard University, 20 Garden Street, Cambridge, MA 02138, USA}

\author[0000-0001-6952-2147]{George~N.~Wong}
\affiliation{School of Natural Sciences, Institute for Advanced Study, 1 Einstein Drive, Princeton, NJ 08540, USA}
\affiliation{Princeton Gravity Initiative, Jadwin Hall, Princeton University, Princeton, NJ 08544, USA}

\author[0000-0003-2966-6220]{Andrew~A.~Chael}
\affiliation{Princeton Gravity Initiative, Jadwin Hall, Princeton University, Princeton, NJ 08544, USA}

\author[0000-0002-4120-3029]{Michael~D.~Johnson}
\affiliation{Center for Astrophysics $\vert$ Harvard \& Smithsonian, 60 Garden Street, Cambridge, MA 02138, USA}
\affiliation{Black Hole Initiative at Harvard University, 20 Garden Street, Cambridge, MA 02138, USA}

\begin{abstract}
    Images of supermassive black hole accretion flows contain features of both curved spacetime and plasma structure. Inferring properties of the spacetime from images requires modeling the plasma properties, and vice versa. The Event Horizon Telescope Collaboration has imaged near-horizon millimeter emission from both Messier 87* (\m{}) and Sagittarius A* (\s{}) with very-long-baseline interferometry (VLBI) and has found a preference for magnetically arrested disk (MAD) accretion in each case. MAD accretion enables spacetime measurements through future observations of the photon ring, the image feature composed of near-orbiting photons. The ordered fields and relatively weak Faraday rotation of MADs yield rotationally symmetric polarization when viewed at modest inclination. In this letter, we utilize this symmetry along with parallel transport symmetries to construct a gain-robust interferometric quantity that detects the transition between the weakly lensed accretion flow image and the strongly lensed photon ring. We predict a shift in polarimetric phases on long baselines and demonstrate that the photon rings in \m{} and \s{} can be unambiguously detected {with sensitive, long-baseline measurements. 
    For \m{}, we find that photon ring detection in snapshot observations requires $\sim1$~mJy sensitivity on $>15$ G$\lambda$ baselines at 230 GHz and above, which could be achieved with space-VLBI or higher-frequency ground-based VLBI. 
    For \s{}, we find that interstellar scattering inhibits photon ring detectability at 230~GHz, but $\sim10$~mJy sensitivity on $>12$ G$\lambda$ baselines at 345~GHz is sufficient, which is accessible from the ground. For both sources, these sensitivity requirements may be relaxed by repeated observations and averaging.}
\end{abstract}

\section{Introduction}

Black holes impose symmetries upon emission near their horizons. In resolved images of optically thin supermassive black hole accretion flows, these symmetries manifest themselves most clearly in the photon ring, the combination of images formed by photon trajectories that reach the observer after half-orbiting the black hole at least once. The photon ring contains self-similar structure indexed by 
the integer $n$, which corresponds to the number of half-orbits 
undertaken by photons before reaching the observer. With each successive half-orbit, the image of the surrounding accretion flow is demagnified exponentially 
along the radial direction into each $n^{\rm th}$ sub-image, asymptotically approaching the ``critical curve,'' a shape predicted by general relativity and determined solely by the mass-to-distance ratio, spin, and viewing inclination of the black hole system \citep{Bardeen_1973, Luminet_1979,Johannsen_2010, Gralla_2019_photonrings, Johnson_2020}. This curve forms the boundary between observer-reaching geodesics that intersect the horizon, and those that do not.

\citet{Himwich_2020} identified additional photon ring symmetries in polarized emission, most relevantly that the Penrose-Walker constant \citep{Walker_Penrose_1970} complex conjugates across adjacent sub-images for axisymmetric emission at large $n$. Recent work has demonstrated analytically that this symmetry corresponds directly to reflections of the electric vector position angle (EVPA) across the origin in face-on images of optically thin, axisymmetric accretion flows around Schwarzschild black holes \citep[][hereafter P22]{Palumbo_2022}. Decompositions of these polarized images into azimuthal modes show complex conjugation in the rotationally symmetric $\beta_2$ mode, which traces near-horizon magnetic field geometry through polarized synchrotron emission \citep{PWP_2020, Emami_2022}. {This formalism was used to measure the accretion state of \m{} in \citet{PaperVII} and \citet{PaperVIII}, hereafter \citetalias{PaperVII} and \citetalias{PaperVIII}.}

In more general models, such as spinning black holes viewed at higher inclination, these symmetries are more approximate. Nonetheless, \citetalias{Palumbo_2022} showed that for magnetically arrested disks (MADs), the class of simulations preferred for Messier 87* (\m{}) and Sagittarius A* (\s{}) by Event Horizon Telescope (EHT) observations, general relativistic magnetohydrodynamic (GRMHD) simulations showed approximate complex conjugation of $\beta_2$ between the primary ($n=0$) and secondary ($n=1$) image. This effect leads to a depolarization of the photon ring region in images which have spiraling polarization, as first noted by \citet{Jimenez_2018}. Thus, using polarimetric observations that resolve out diffuse $n=0$ structure and are dominated by $n=1$ emission, the existence of the photon ring can be demonstrated by measuring the $\beta_2$ phase of the $n=1$ image.

\begin{figure*}[htbp]
    \centering
    \includegraphics[width=\textwidth]{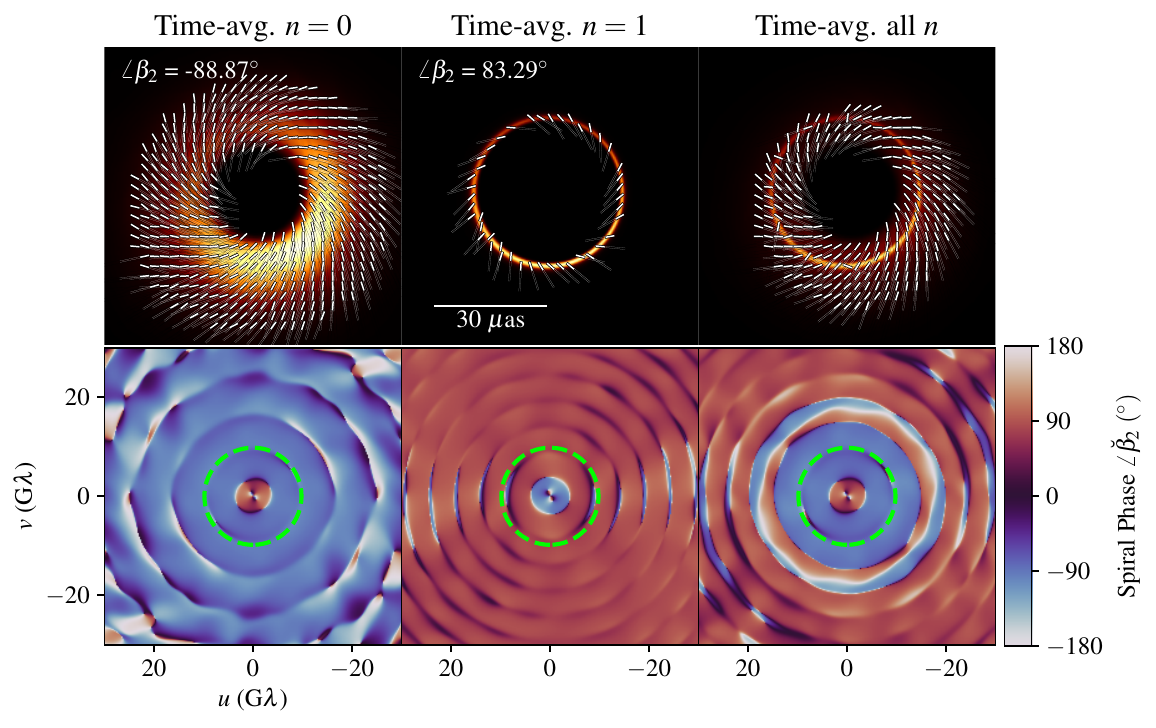}
    \caption{Spiral quotient phase (bottom row) for a time-averaged MAD simulation with spin $a_*=0.5$ and $R_{\rm high}=80$, ray traced at 230 GHz and decomposed into the direct ($n=0$), indirect ($n=1$) and full (all $n$) images (top row). In the bottom right panel, the $n=0$ phase dominates on short baselines until the direct image thickness is resolved, at which point the $n=1$ phase dominates the combined signal. The dashed green circle shows Earth-diameter baselines, which never reach the $n=1$-dominated regime at 230 GHz.
    }
    \label{fig:230decomp}
\end{figure*}

In this letter, we construct a polarized interferometric quantity that is invariant under source translations and robust to unknown gain amplitudes and phases. This quantity serves as a Fourier analogue of $\beta_2$ and refines a similar construction carried out in \citetalias{PaperVIII}. We compute this quantity on Earth-based and Earth-space very-long-baseline interferometry (VLBI) baselines for simulated images and find a detectable photon ring signature in the phase of the interferometric quantity $\breve{\beta}_2$. {Throughout, we use $\angle$ to denote phases of complex numbers. }

We develop the observable in \autoref{sec:vlbi}. We treat observational considerations in simulations of \m{} and \s{} in \autoref{sec:obs}. We conclude with a discussion in \autoref{sec:discussion}. We diagram some mathematical details in \autoref{sec:bessel} and analyze potential Earth-based observing sites in \autoref{sec:sites}.

\section{Interferometric Polarization Spirals}
\label{sec:vlbi}

The image-domain definition of the $\beta_m$ decomposition radially averages azimuthal structure in linear polarization in terms of the usual Stokes parameters $Q$ and $U$.  {We consider images in polar coordinates $r$ and $\phi$, with $\phi$ increasing east of north and right ascension increasing east, to the left, as usual.  As in \citet{PWP_2020}, for an image with complex polarization scalar $P(r,\phi)=Q(r,\phi)+i U(r,\phi)$}, each modal coefficient is given by
\begin{align}
\label{eq:imbetam}
    \beta_m &= \dfrac{1}{I_{\rm tot}} \int\limits_{0}^{\infty} \int\limits_0^{2 \pi} P(r, \phi) \, e^{- i m \phi} \; r \mathop{dr}\mathop{d\phi}  .
\end{align}
Here, the upper bound on the integral in $r$ is set effectively by the field of view of an observed image. The $\beta_2$ coefficient provides an image-averaged measurement of the rotational symmetry of the observed EVPA. 

We now construct interferometric $\beta_2$ modes, which were first described in Appendix A of \citetalias{PaperVIII}. A perfectly calibrated interferometer measures components of the Fourier transforms of each Stokes parameter image, $\tilde{I}$, $\tilde{Q}$, and $\tilde{U}$. Following \citet{Kamionkowski_2016}, we may rotate into an interferometric $E$ and $B$ mode basis by applying a rotation by twice the angle of each measured visibility $\tilde{Q}(u,v)$ and $\tilde{U}(u,v)$, where $u$ and $v$ are two-dimensional Fourier coefficients conjugate to right ascension and declination, respectively \citep{TMS}.  The rotation yields quantities $\tilde{E}$ and $\tilde{B}$:
\begin{gather}
\label{eq:rot}
 \begin{bmatrix} \tilde{E}(\rho,\theta) \\ \tilde{B}(\rho,\theta) \end{bmatrix}
 =
  \begin{bmatrix}
   \cos 2 \theta &
   \sin 2 \theta \\
   -\sin 2 \theta &
   \cos 2 \theta 
   \end{bmatrix}
   \begin{bmatrix}
   \tilde{Q}(\rho,\theta) \\ \tilde{U}(\rho,\theta)
   \end{bmatrix},
\end{gather}
where $\rho$ is the coordinate radius and $\theta \equiv \arctan(u/v)$ is the position angle measured east of north of each point $(u,v)$.\footnote{ At this point, the construction in Appendix A of \citetalias{PaperVIII} would normalize each of $\tilde{E}$ and $\tilde{B}$ by $|\tilde{I}|$ {(Equation A13 in that work)} and construct a quantity encoding EVPA spiral phase with a reliance on knowledge of the image center (and therefore absolute phase calibration of interferometric visibilities). {Subsection A.3 of that work} connects the signs of these interferometric modes to the image-domain $\beta_2$ phase; the precise connection {(and sign convention in Equation A14)} claimed in the paper holds only for short baselines, whereas for general baseline lengths, additional signs are imposed by image structure (for example, when the Bessel function response to a ring-like structure passes through a null and changes sign, as shown in the orange curve in the left panel of \autoref{fig:bessel}).}  

\begin{figure*}[htbp]
    \centering
    \includegraphics[width=\textwidth]{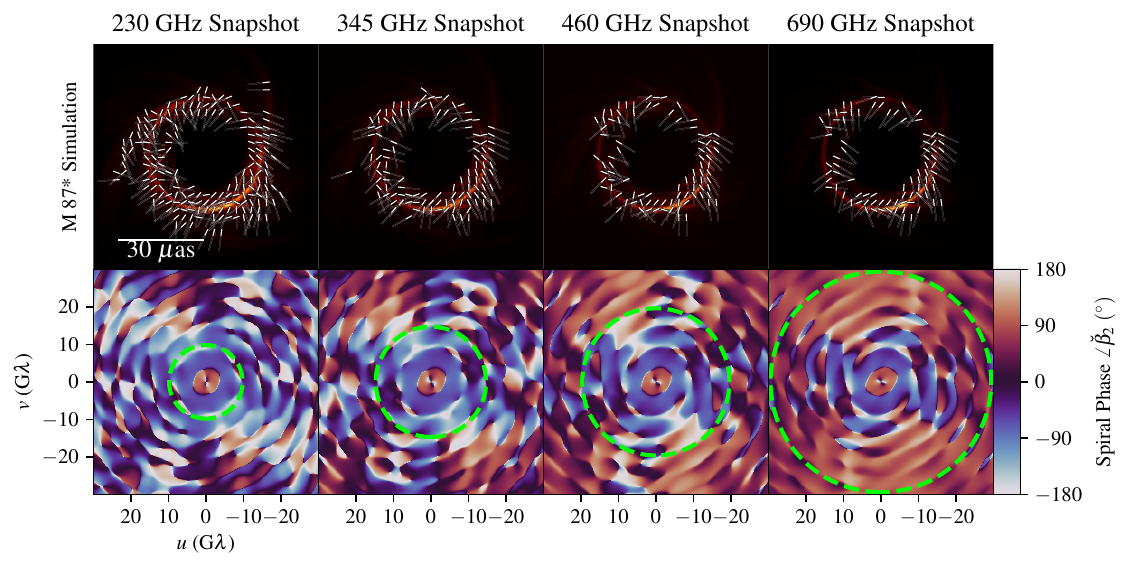}
    \caption{Spiral quotient phase (bottom row) for snapshots of the same GRMHD fluid frame at four different frequencies, ray traced with \m{} parameters. The dashed green circle shows Earth-diameter baselines at each frequency. At higher frequencies, the direct image becomes relatively fainter compared to the indirect image, causing the crossing point between $n=0$ and $n=1$ to move inwards to shorter baselines just in reach of Earth-based VLBI; at 345 GHz, Earth-space VLBI appears viable for finding the photon ring phase change.}
    \label{fig:snaps}
\end{figure*}

In order to produce a quantity that captures rotationally symmetric polarization structure in ring-like images while avoiding both the need for phase calibration and the sign structure imposed by oscillating Fourier signatures of image features, we construct the following polarimetric quantities:
\begin{align}
    \breve{e}(u,v) &\equiv \frac{\tilde{E}(u,v)}{\tilde{I}(u,v)},\\
    \breve{b}(u,v) &\equiv \frac{\tilde{B}(u,v)}{\tilde{I}(u,v)},\\
    \breve{\beta}_2(u,v) &\equiv{\rm Re}(\breve{e}(u,v)) + i \, {\rm Re}(\breve{b}(u,v)).
    \label{eq:beta2}
\end{align}
Dividing by $\tilde{I}$ removes most unknown gain contributions (in both amplitude and phase) from the signal, {except for unknown (but typically stable and calibrated) right-left gain ratios and leakage terms as shown in \autoref{sec:rime}. This construction also makes $\breve{e}$, $\breve{b}$, and $\breve{\beta}_2$ close cousins to the polarimetric ratio or interferometric fractional polarization $\breve{m}$ and other cross-hand-to-parallel-hand polarization ratios, which share similar gain robustness \citep{Roberts_1994, Johnson_2014}.} 

{In addition, these ratios are insensitive to some classes of physical corruptions, such as are introduced by scattering in the ionized interstellar medium. Specifically, the dominant effects of scattering are a convolution with a ``diffractive'' blurring kernel. Because the interstellar medium is not appreciably birefringent at mm wavelengths, the Fourier manifestation of interstellar scattering cancels in the formation of a quotient between $\tilde{Q}$ or $\tilde{U}$ and $\tilde{I}$, though the signal-to-noise ratio is depressed by the diffractive kernel. 
On long baselines, additional effects of ``refractive'' scattering become significant, which cannot be described as an image convolution. On these baselines, we will demonstrate that refractive scattering contaminates the visibility quotients.} 


In constructing $\breve{\beta}_2$, we take the real part of each of $\breve{e}$ and $\breve{b}$ because the real parts of $\tilde{E}$ and $\tilde{B}$ project out rotationally symmetric structure corresponding to even $\beta_m$ modes. This relation stems from basic properties of the Fourier transforms of even and odd functions, and is apparent from Equation A10 of \citetalias{PaperVIII}, reproduced here:
\begin{align}
    \tilde{E}(\rho,\theta) &= \sum\limits_{m=-\infty}^{\infty}i^{-m}{\rm Re}[\beta_m e^{i (m-2)\theta}F_m(\rho)],\nonumber\\
    \tilde{B}(\rho,\theta) &= \sum\limits_{m=-\infty}^{\infty}i^{-m}{\rm Im}[\beta_m e^{i (m-2)\theta}F_m(\rho)].
\end{align}
Here, $F_m(\rho)$ is the Hankel transform of a radial function $f_m(r)$ describing radial image structure associated with sinusoidal variation of order $m$:
\begin{align}
    F_m(\rho) = \int_0^\infty f_m(r)J_m(2 \pi r \rho) r dr,
\end{align}
with $J_m$ as the Bessel function of the first kind of order $m$.
The symmetry projection property  does not formally generalize to to $\breve{e}$ and $\breve{b}$ for arbitrary $\tilde{I}$, but the quotient preserves this property for images in which the image features in $I$ and $P$ are similar.

The spiral quotient $\breve{\beta}_2(u,v)$ thus encodes a translation-invariant notion of rotationally symmetric polarization structure. It is notable that an interferometric measure of rotational symmetry can be constructed without a defined image center; the situation is analogous to the definition of the image second moment in \citet{Issaoun_2019}, where image covariance about an unspecified center of light is used. Ultimately, the phase of the quantity encodes the dominant spiral phase at a particular spatial frequency given by $(u,v)$; if the photon ring reversal is to be observed, there must be an observable reversal in this phase across $(u,v)$-distance $\rho$.

We {now} examine the detailed character of $\breve{\beta}_2$ by considering a concrete example. 

\subsection{Thin Polarized Rings}
As discussed in \citet{PWP_2020}, we consider a thin ring with total flux {$I_{\rm tot}$} and diameter $d$  in radians with rotationally symmetric polarization; the total intensity and polarization in image polar coordinates $r$ and $\phi$ are given by
\begin{align}
    I(r,\phi) &= \frac{I_{\rm tot}}{\pi d}\delta\left(r-\frac{d}{2}\right),\\
    P(r,\phi) &= \beta_2  \frac{I_{\rm tot}}{\pi d}\delta\left(r-\frac{d}{2}\right)e^{i 2 \phi},
\end{align}
where $\beta_2$ sets the rotationally symmetric EVPA spiral phase. The visibility responses are then
\begin{align}
    \tilde{I}(\rho,\theta) &= I_{\rm tot} J_0(\pi d \rho),\\
    \tilde{P}(\rho,\theta) &= -\beta_2  I_{\rm tot} J_2(\pi d \rho) e^{i 2 \theta}.
\end{align}
We then project into $\tilde{Q}$ and $\tilde{U}$:
\begin{align}
    \tilde{Q}(\rho,\theta) &= {\rm Re}(\tilde{P}(\rho,\theta)),\nonumber\\
    &= -I_{\rm tot} J_2(\pi d \rho)\\ \nonumber &\times \left[{\rm Re}(\beta_2) \cos2\theta - {\rm Im}(\beta_2) \sin2\theta\right],\\
    \tilde{U}(\rho,\theta) &= {\rm Im}(\tilde{P}(\rho,\theta)),\nonumber\\
    &= -I_{\rm tot} J_2(\pi d \rho)\\ \nonumber &\times \left[{\rm Re}(\beta_2) \sin2\theta + {\rm Im}(\beta_2) \cos2\theta\right].
\end{align}
$\tilde{E}$ and $\tilde{B}$ project out real and imaginary parts of $\beta_2$ as expected, using \autoref{eq:rot}:
\begin{align}
    \tilde{E}(\rho,\theta) &= -I_{\rm tot} {\rm Re}(\beta_2) J_2(\pi d u),\\
    \tilde{B}(\rho,\theta) &= -I_{\rm tot} {\rm Im}(\beta_2) J_2(\pi d u).
\end{align}
Lastly, we divide by $\tilde{I}$ and sum the two quotients to finish the $\breve{\beta}_2$ construction:
\begin{align}
    \breve{\beta}_2 &= -\left[{\rm Re}(\beta_2)+i {\rm Im}(\beta_2)\right] \frac{J_2(\pi d u)}{J_0 (\pi d u)},\nonumber\\
    & = -\beta_2  \frac{J_2(\pi d u)}{J_0 (\pi d u)}.
\end{align}
As we discuss at length in \autoref{sec:bessel}, the ratio of $J_2$ to $J_0$ has a nearly constant negative sign after the first null of $J_0$ \citep{McMahon_1894}, meaning
\begin{align}
    \angle{\breve{\beta}_2} &\approx \angle \beta_2.
\end{align}

\subsection{General Images and Simulations}

We now examine $\angle\breve{\beta}_2$ for baseline lengths of interest for VLBI in an example GRMHD simulation which is in decent agreement with the EHT Collaboration's constraints on both \m{} and \s{}, though with slight differences in viewing inclination and electron distribution function (see \citetalias{PaperVIII} and \citealt{SgrA_PaperV}). This simulation is of a magnetically arrested disk with dimensionless black hole spin $a_*=0.5$ spinning prograde with respect to the large scale accretion flow. The electron distribution function post-processing parameter $R_{\rm high}=80$ \citep[see][ for details]{Mosci_2016,PaperV}. This simulation is generally typical by the sub-image polarization standards of \citetalias{Palumbo_2022}; its subimage is polarized and demonstrates a near complex conjugation of $\beta_2$ across sub-image index.  

We ray trace images of this simulation at 230, 345, 460, and 690 GHz, with 480 pixels across 160 $\mu$as on each edge at each frequency except 690 GHz, where 600 pixels are used to robustly capture fine structures sampled by 690 GHz Earth-diameter baselines. The 480 pixel images have one third of a $\mu$as per pixel, corresponding to a spatial frequency of $\sim 600$ G$\lambda$; phases of Fourier quantities are thus highly robust in the at-most $50\sqrt{2}$ G$\lambda$ coefficients we compute. The GRMHD simulations were done with \texttt{iharm3D} \citep{Gammie_HARM_2003, IHARM3d_prather}; ray tracing was performed using \texttt{ipole} \citep{IPOLE_2018}. Additional details on the image generation process may be found in \citet{Wong_2022}. 

We ray trace the fluid snapshots from this simulation with two sets of parameters befitting \m{} and \s{}. For \m{}, we use the mass-to-distance ratio measured by the EHTC (corresponding to the ray traced and decomposed images in \citetalias{Palumbo_2022}) with a $163^\circ$ inclination (such that fluid motion is clockwise on the sky) and we rotate images so that the approaching jet is oriented $288^\circ$ degrees east of north. For \s{}, we use the mass-to-distance ratio in \citet{SgrA_PaperV} and a $150^\circ$ inclination with accretion flowing clockwise in the sky; we do not rotate images, so the approaching outflow is oriented towards the viewer and upwards on the sky by $30^\circ$.

\begin{figure*}
    \centering
    \includegraphics[width=\textwidth]{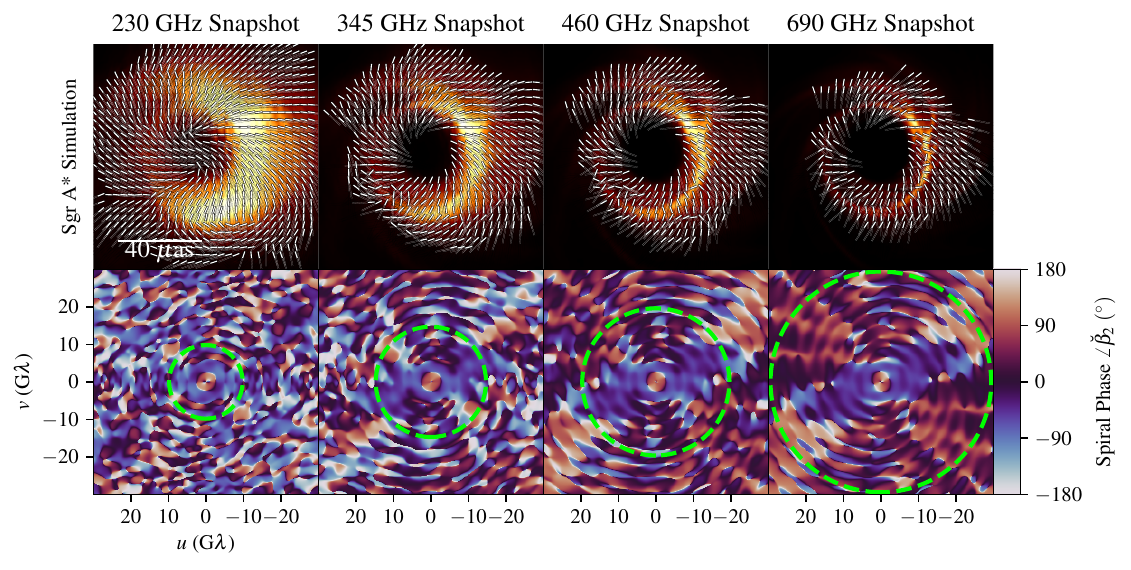}
    \caption{Spiral quotient phase for scattered snapshots of the same \s{} GRMHD fluid frame at 4 frequencies. The dashed green circle shows Earth-diameter baselines at each frequency.
    Images are scattered by the same (frequency-dependent) scattering screen looking towards the galactic center. Refractive scattering distributes diffuse polarization structure into higher spatial frequencies, outshining the photon ring phase at 230 GHz.}
    \label{fig:sgra_snap}
\end{figure*}
\autoref{fig:230decomp} shows the time-averaged 230 GHz image of the \m{} simulation decomposed into its direct ($n=0$), indirect ($n=1$), and full (all $n$) images, as well as the corresponding spiral phase $\angle\breve{\beta}_2$ over the interval from $-30$ to $30$ G$\lambda$ in $u$ and $v$. We see that the phase corresponds to the image-domain $\beta_2$ in each of the individual $n$ cases, but in the full image, there is a transition between the dominance of $n=0$ and $n=1$ at a particular radius in the $(u,v)$ plane. This radius is sensitive to the relative size and brightness of $n=0$ and $n=1$ images; time-averaging drastically reduces the fine structure in the $n=0$ image, causing a clear cutoff in the bottom right panel as the $n=0$ thickness is resolved.

Snapshots, however, introduce significant fine structure to the $n=0$ image, as shown for all four frequencies of interest for \m{} in \autoref{fig:snaps} and for \s{} in \autoref{fig:sgra_snap}. For \s{}, we scatter images using a single realization of the frequency-dependent scattering screen implemented by \citet{Johnson_2016} in the {\tt stochastic-optics} library of {\tt eht-imaging}. In both \m{} and \s{}, the phase transition between $n=0$ and $n=1$ shifts inward over frequency primarily because the brightness ratio between the direct and indirect image shifts to favor the indirect image at higher frequencies, likely due to optical depth effects. In \s{}, at lower frequencies, the refractive noise from scattering mixes the diffuse $n=0$ structure into finer spatial scales, outshining the photon ring on long baselines.

\section{Observational Prospects}
\label{sec:obs}

We examine the practicality of observing the spiral quotient phase change caused by the photon ring by considering baseline lengths corresponding to Earth-based VLBI at each frequency and Earth-space VLBI at 230 and 345 GHz. We examine this problem in general terms by computing properties along the $u$ and $v$ axes in the Fourier domain; in \autoref{sec:sites}, we consider the sampling and temporal statistics of particular sites of interest for \m{} and \s{}.

\begin{figure*}[ht!]
    \centering
    \includegraphics[width=\textwidth]{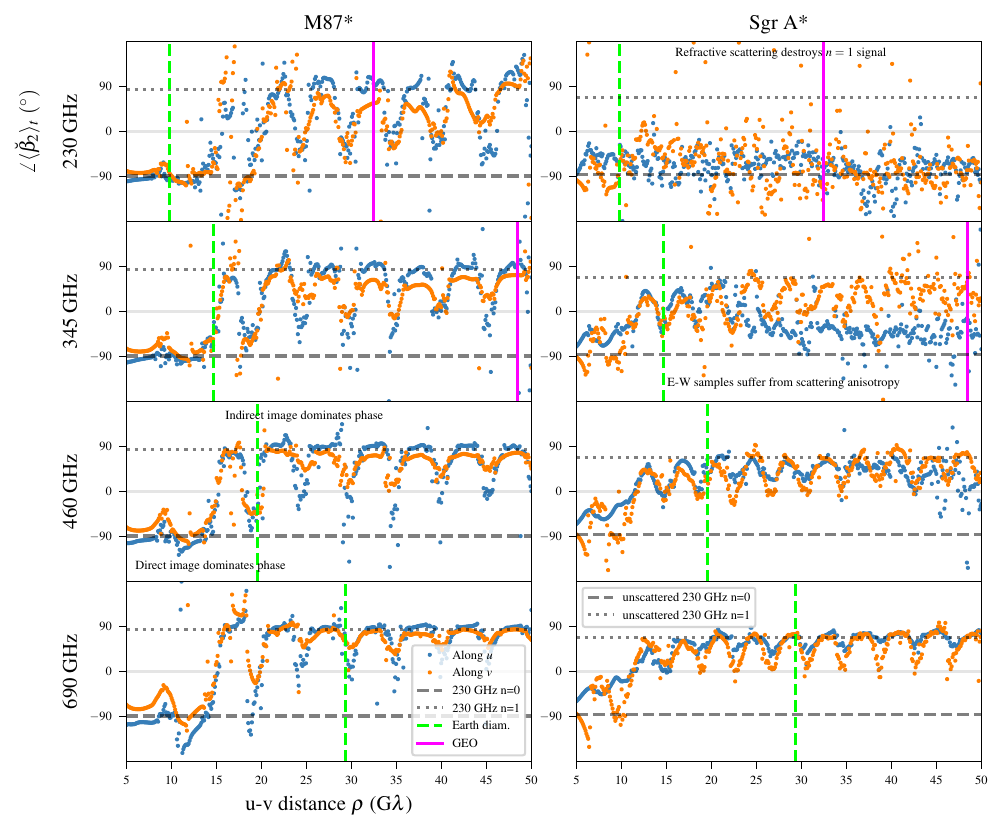}
    \caption{Phases of the time-averaged $\breve{\beta}_2$ for the \m{} and \s{} simulations sampled along the $u$ and $v$ axes in the Fourier domain, along with the time-averaged image-domain 230 GHz decomposed phases from the simulation, unscattered in the \s{} case. In the \m{} simulation, the $u$ axis is nearly aligned with the jet, whereas in the \s{} simulation, $u$ is aligned with the line of nodes, so any ``jet'' would be aligned with $v$. Observations of each snapshot are computed {without thermal noise assuming known right-left gain ratios and leakage terms} albeit with unknown absolute amplitude or phase calibration; values of $\breve{e}$ and $\breve{b}$ are then each averaged to produce the mean $\breve{\beta}_2$. Increasing frequency is most impactful in pushing the indirect image-dominated regime to the left, near 15 G$\lambda$ at 690 GHz. Dashed green lines show the Earth diameter baseline at each frequency, while solid magenta lines show the geostationary orbital radius, which is dominated by the indirect image at all frequencies after averaging. The direct and indirect image values $\angle\breve{\beta}_2$ are each nearly achromatic, suggesting that the emission region does not change significantly over frequency.}
    \label{fig:time}
\end{figure*}

\subsection{Intrinsic Averages of the Spiral Quotient}

\begin{figure*}[t!]
    \centering
    \includegraphics[width=\textwidth]{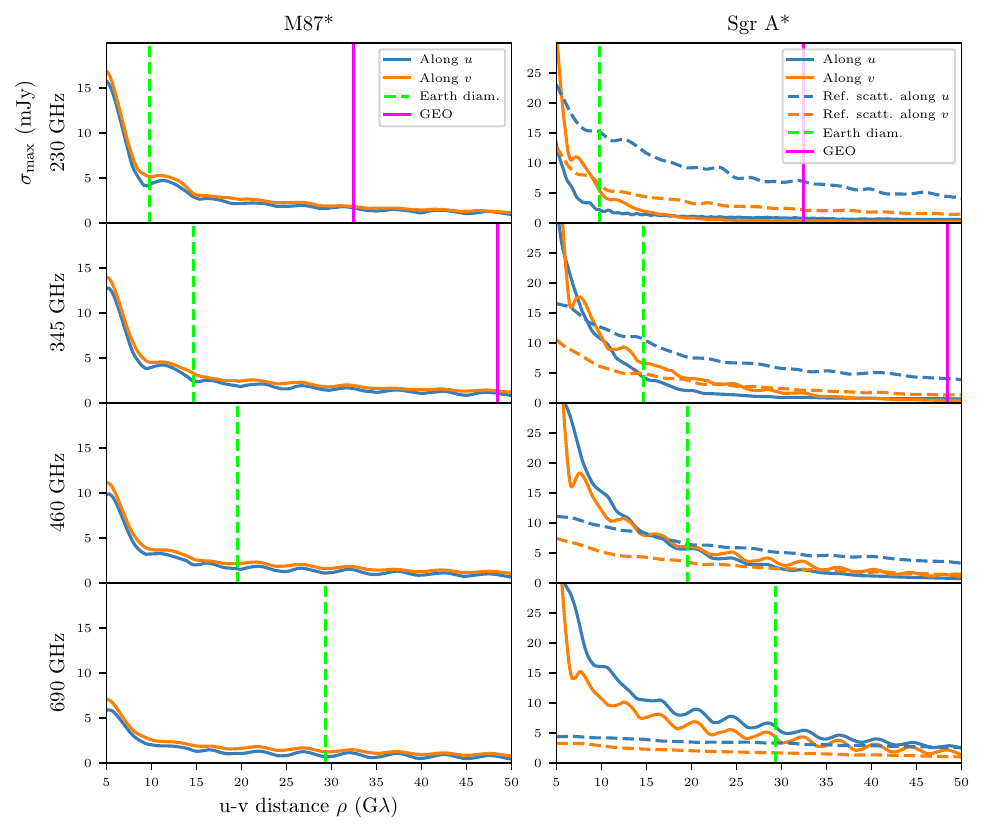}
    \caption{Maximum permissible thermal noise on individual visibility measurements allowed for robust spiral quotient phase signal in {snapshot} detections as given by \autoref{eq:noise}, averaged at each point over one thousand gravitational times for \m{} (left) and \s{} (right), plotted alongside the refractive scattering noise on each axis in dashed lines for \s{}. Differences between sampling along $u$ and $v$ reflect intrinsic source structure caused by non-zero inclination, as well as scattering anisotropy in the case of \s{}. The sensitivity required to recover the phase in \s{} with Earth-diameter interferometry (lime green) stays approximately constant in frequency, though the phase is dominated by refractive sub-structure at the lower frequencies.}
    \label{fig:noise}
\end{figure*}

Given the phase structure in \autoref{fig:snaps} and \autoref{fig:sgra_snap}, one may worry that an individual observation on a small number of baselines could mislead conclusions about photon ring existence due to transient image structures. It would thus be useful to be able to average the complex quantities $\breve{e}$ and $\breve{b}$ over many epochs and extract the average phase of $\breve{\beta}_2$. One would expect that as the direct image of the flow varies in thickness and diameter slightly over time, the nulls in its visibility response may slide inward and outward leading to regions of unpredictable phase; the success of an experiment searching for a consistent phase offset between the direct and indirect emission would thus rely on having coverage far enough from the nulls to have stable spiral phase and non-zero spiral amplitude on average.

\autoref{fig:time} shows the result of measuring $\breve{e}$ and $\breve{b}$ at each of the four frequencies on baselines along $u$ and $v$ for each snapshot of the \m{} and \s{} simulations, averaging the results over time, and computing the average $\angle\breve{\beta}_2$. Averages are computed over 5000 gravitational times ($G M / c^3$) in the simulation as ray traced for \m{} and \s{}, where $G$ is the gravitational constant, $M$ is the black hole mass, and $c$ is the speed of light. In \s{}, we use a random realization of the scattering screen in each frame of the simulation, effectively averaging over both intrinsic and extrinsic structure to produce the phase curves. No instrumental noise is added; the averaging thus represents only an average over variation on the sky, {assuming that right-left gain ratios and D-terms are known, while absolute amplitude and phase calibration is unnecessary.}

We observe that in \m{}, at low frequencies, the transition to photon ring domination is muddled by source variation and brighter $n=0$ structure which becomes more crisp as frequency increases. In \s{}, we observe that the refractive scattering noise imposes small scale structure at low frequencies that obscures the photon ring on long baselines (with notable differences between $u$ and $v$ at 345 GHz due to the stronger scattering on the E-W axis). However, at 460 GHz and higher, the scattering is nearly negligible. It is noteworthy that the phase in the scattering-dominated regimes of \s{} is not random, but instead indicative of the $n=0$ structure on average; this property again suggests that the refractive scattering mixes diffuse structure into finer angular scales, as is most prominent in the bottom left of \autoref{fig:sgra_snap}.

{In summary, using these polarimetric visibility quotients the photon ring in \m{} can only be detected with VLBI observations at 460~GHz and higher on the ground or at 230~GHz and higher with space-VLBI baselines.} {For \s{}, the photon ring never dominates the time-averaged signal at 230~GHz due to scattering corruptions. However, at 345~GHz, the photon ring dominates near-Earth-diameter baselines (particular N-S baselines, which are the most weakly affected by scattering), though longer baselines (particularly E-W) baselines are corrupted by refractive scattering. At 460 and~690 GHz, the effects of refractive scattering are negligible, and the photon ring phase transition is apparent, even on Earth baselines.} 
{In each source, these simulations also show transition regions (approximately 15 G$\lambda$ for \m{} and 12 G$\lambda$ for Sgr A*) beyond which the $n=1$ ring begins to dominate the observed polarization. The exact value of this transition point will depend on simulation parameters, but intuitively, the larger angular size of \s{} suggests a transition at a smaller baseline length.}

\subsection{Sensitivity Requirements}

Error propagates into $\breve{\beta}_2$ similarly to $\Breve{m}$, the interferometric fractional polarization (see appendices A and B of \citealt{Chael_2016}) in the high signal-to-noise ratio limit:
\begin{align}
    \label{eq:singlenoise}
    \sigma_{\breve{\beta}_2} &\approx \sigma \sqrt{\frac{2}{|\tilde{I}|^2} + \frac{|\tilde{P}|^2}{|\tilde{I}|^4}},\\
    \sigma_{\angle\breve{\beta}_2} &\approx \frac{\sigma_{\breve{\beta}_2}}{|\breve{\beta}_2|}.
\end{align}
Here, once again $\tilde{P} = \tilde{Q}+i\tilde{U}$ and $\sigma$ is the thermal noise on individual baseline amplitude measurements, assumed to be equal for $\tilde{I}$, $\tilde{Q}$, and $\tilde{U}$. The amplitude of the spiral quotient $|\breve{\beta}_2|$ is proportional to the fractional polarization, so unsurprisingly, the phase error decreases as the fractional polarization increases.

As shown in Figure 2 of \citetalias{Palumbo_2022}, the direct and indirect image $\beta_2$ phases are typically separated by at least $\pi/2$ radians in the magnetically arrested disks best suited for fitting observations of \m{}. To have a clear photon ring signal in the spiral quotient phase, we therefore choose a target phase uncertainty of $\sigma_{\angle\breve{\beta}_2} \leq \pi/4$. This condition yields a constraint on thermal noise that can be calibrated to simulations:
\begin{align}
    \sigma_{\rm max} &\leq \frac{\pi |\breve{\beta}_2|}{4\sqrt{\frac{2}{|\tilde{I}|^2}+\frac{|\tilde{P}|^2}{|\tilde{I}|^4}}}.
    \label{eq:noise}
\end{align}

In order to inform future hardware requirements, as well as to broadly determine whether this signature will ever be detectable from the Earth's surface, we now compute \autoref{eq:noise} for each GRMHD snapshot used to produce \autoref{fig:time}. Once again, the values of each visibility used to evaluate the target noise level are {computed without the addition of thermal noise, assuming known right-left gain ratios and leakage terms}. In the case of \s{}, we generate 2000 realizations of the scattered GRMHD images to estimate the root-mean-squared variation in complex visibilities as a result of scattering (``refractive noise'') along east-west and north-south baselines.

As shown in \autoref{fig:noise}, for the example simulation of \m{}, a thermal noise on individual visibility measurements of $\sim 2$ mJy will be required to capture the 460 or 690 GHz spiral phase in a single observation of \m{}, while in \s{} 10 mJy will be required. In \s{}, even with a perfect noiseless instrument, on long baselines at low frequencies, the interferometric signal is dominated by refractive sub-structure. Comparing to \autoref{fig:time}, the regions in sampling of \s{} in which the phase departs from the expected $n=1$ phase correspond to where the refractive scattering curves (dashed lines) exceed the intrinsic signal strength.

However, the photon ring detection need not be captured in a single observation. The translation invariance and gain robustness of $\breve{\beta}_2$ allow coherent averaging over multiple observations, potentially spanning years of accretion, ultimately yielding values that correspond to an instrument-corrupted sub-sampling of \autoref{fig:time}. Assuming the accretion state of \m{} is approximately constant or has a typical quiescent magnetic field structure, even observations with signal-to-noise ratio less than 1 may be useful in constraining the sub-image relation; the only requirement for a visibility to be useful is then a detection in Stokes $I$, which is less stringent.

It must be noted that we have chosen only a single simulation for this test, and that this simulation was ray traced to produce an average flux density of 0.5\,Jy for \m{} and 2.5 Jy for \s{}, from which individual realizations of the flow may differ significantly. Moreover, the thermal noise requirement is sensitive to the relative brightness of the photon ring compared to the direct emission, and this flux ratio can vary by factors of two across different GRMHD parameters, and factors of a few during flares \citep[see, e.g.][]{Gelles_2022, Wielgus_2022}. Ultimately, expectations will need to be refined as next-generation hardware capabilities are better understood, and as EHT observations elucidate the typical brightness of fine structure in \m{} and \s{}.

\section{Discussion}
\label{sec:discussion}

In this letter, we have constructed a gain-robust, image translation-invariant interferometric observable $\breve{\beta}_2$ which is intrinsically sensitive to rotationally symmetric polarization structures. In magnetically arrested disks (the increasingly favored accretion paradigm for low luminosity active galactic nuclei), this observable elegantly captures the transition between direct image and photon ring-dominated emission on long interferometric baselines. We have shown that VLBI at high frequencies is capable of unambiguously detecting the photon ring with this observable in \m{} and \s{}; however, we estimate that the required sensitivities for snapshot detection are stringent, at the level of {$\leq10$ mJy.}

These sensitivity challenges could potentially be addressed in many ways. The gain-robustness and translational invariance of $\breve{\beta}_2$ permits the averaging of many epochs, meaning that the primary concern at high frequencies is obtaining detections in Stokes $I$. Long integration times may enable these detections even in sub-optimal weather; these integrations may be enabled by frequency phase transfer from 230 GHz, though not all sites can support the prerequisite simultaneous multifrequency observation \citep[see, e.g.][]{RD_Multiview, RD_review, RD_ngeht}. Moving to even larger bandwidths will also help overcome these sensitivity limitations, as the prominent features of the direct and indirect image persist over wide radial ranges in the $(u,v)$ plane permitting large bandwidth smearing.

Though we have restricted ourselves to single-baseline measurements, it is conceivable that interferometric sensitivity to the photon ring is enough to depolarize reconstructed images from future EHT data even at 345 GHz. Indeed, \citet{Jimenez_2018} identified frequent depolarization of the photon ring image region that remains apparent even with small levels of blurring, as might be expected of 345 GHz reconstructions. Moreover, in imaging or hybrid imaging-modeling experiments that attempt to extract the photon ring \citep[see, e.g.][]{Broderick_2022_ph, Lockhart_2022, Tiede_2022}, our approach can build confidence in extracted photon rings by demonstrating before model fitting that a sharp, oppositely polarized feature is present.

As discussed at length in \citet{Palumbo_2022} and treated analytically in \citet{Himwich_2020}, the relationship between the direct and indirect image $\angle\beta_2$ is a function of the black hole spin, the magnetic field in the emitting region, and the geometry of the emission. At first, a detection of a large, persistent difference in the polarization spiral phase between the intermediate and extremely long baselines is a detection of the photon ring that depends only on broad belief that the emission is optically thin. However, by specifying and model-fitting the emission, such as in semi-analytic models or by making stronger GRMHD assumptions, the spin and mean magnetic field morphology may be tightly constrained by the quiescent $\breve{\beta}_2$ phases. In the context of proposed demographic studies of supermassive black holes with VLBI \citep{Pesce_2021, Pesce_2022}, observational schemes such as ours may enable single-baseline spin and photon ring demographic studies with fewer model assumptions than other approaches. 

The interferometry of the future may well target the $n=2$ image for its excellent approximation to the critical curve; our construction predicts an infinite sequence of phase transitions in $\angle\breve{\beta}_2$ as each subsequent $n$ becomes dominant, provided that optical and Faraday rotation depths internal to the accretion flow permit structured polarization on long photon trajectories. Though the polarization of the $n=2$ image in GRMHD simulations has not been studied exhaustively, the polarization spiral will likely resemble the $n=0$ image spiral in magnetically arrested disks.

Though we have used polarization quotients such as $\breve{e} = \tilde{E} / \tilde{I}$ throughout this letter in order to remove dependencies on the image center as well as unknown gains, these quotients may not serve as optimal estimators in realistic observations. For example, the quantities $\tilde{E}\tilde{I}^*$ and $\tilde{B}\tilde{I}^*$ are also translation invariant (and thus robust to unknown gain phases). Though these quantities are not robust to unknown gain amplitudes, products typically have more favorable low signal-to-noise ratio statistics than quotients. Ultimately, the aim of this letter is to lay out key ideas of relevance to the VLBI observations of the near future; finding the optimal interferometric data product by which the photon ring will be robustly detected is a task for future work.

\acknowledgments
We thank Lindy Blackburn,  Ramesh Narayan, and Dominic Pesce for many useful conversations. We also thank Jim Moran for thorough comments on the manuscript. We are also very grateful to our reviewer for many thoughtful comments which greatly improved this letter. This work was supported by the Black Hole Initiative at Harvard University, which is funded by grants from the John Templeton Foundation and the Gordon and Betty Moore Foundation to Harvard University (GBMF-5278). D.C.M.P. was supported by National Science Foundation grants AST 19-35980 and AST 20-34306. G.N.W. gratefully acknowledges support from the Taplin Fellowship. AC was supported by the Gravity Initiative at Princeton University.

\pagebreak
\appendix

\section{Radio Inteferometric Corruptions in Polarimetric Quotients}
\restartappendixnumbering
\label{sec:rime}
In this appendix, we review the radio interferometer measurement equation (RIME) outlined for analysis of EHT observations in section 2 of \citetalias{PaperVII}; the results reached in this appendix are analagous to similar arguments in \citet{Roberts_1994} and \citet{Johnson_2014}, and as usual, the most comprehensive primer can be found in \citet{TMS}. This review addresses complications in measuring polarization which are present even in the limit of high signal-to-noise ratio, corresponding to fundamental unknowns (particularly gains) in the interferometric problem which are typically calibrated away. Throughout this section, we distinguish intrinsic variables from measured variables with a prime ($'$).

Nearly all EHT stations measure polarization in a circular basis, sampling the left-handed ($L$) and right-handed ($R$) complex-valued electric field. For a pair of telescopes indicated by the indices $j$ and $k$, the corresponding complex correlation matrix $\rho_{jk}$ is given as follows:
\begin{align}
     \rho_{jk}
     &=
      \begin{bmatrix}
      R_j R_k^* & R_j L_k^* \\
      L_j R_k^* & L_j L_k^*
      \end{bmatrix},\\
      &=
      \begin{bmatrix}
      \tilde{I}_{jk} + \tilde{V}_{jk} & \tilde{Q}_{jk} + i \tilde{U}_{jk} \\
      \tilde{Q}_{jk} - i \tilde{U}_{jk} & \tilde{I}_{jk} - \tilde{V}_{jk}
      \end{bmatrix}.
    \label{eq:rho}
\end{align}
Here, the second line shows the relation between the correlation matrix and Stokes visibilities; though we do not use it in this letter, $\tilde{V}$ refers to the Stokes $V$ visibility.

In practice, the measured correlation matrix $\rho_{jk}'$ is corrupted by complex time-dependent gains, leakages, and field rotations, represented as matrices $\bf G$, $\bf D$, and $\bf\Phi$, which are unique to each station and are joined in a Jones matrix $\bf J$:
\begin{align}
    {\bf J} &= {\bf G D \Phi},\\
    {\bf G} &= \begin{bmatrix}
    G_R & 0 \\ 0 &G_L
    \end{bmatrix},\\
    {\bf D} &= \begin{bmatrix}
    1 & D_R \\ D_L & 1
    \end{bmatrix},\\
    {\bf \Phi} &= \begin{bmatrix}
    e^{-i\Phi} & 0 \\ 0 & e^{i \Phi}
    \end{bmatrix}.
\end{align}
As explained in detail in subsection 3.2 of \citetalias{PaperVII}, the time-dependent field rotation $\Phi$ is a function only of the geometry of each antenna (and its feed orientation) as well as the parallactic angle and elevation of the source over the course of an observation. These are all known a priori to high precision, and so their effects can be removed (except for contributions from leakage, as will be apparent in the expressions that follow) by applying the inverse field rotation matrix $\bf\Phi^\dag$.  

$\bf J$ acts on the measured electric fields at a single station; adding in the geometric derotation matrix for each station, the resulting time-dependent corrupted correlation matrix for a pair of stations is then given by
\begin{align}
    \rho'_{jk} &= {\bf \Phi^\dag_j}{\bf J_j} \rho_{jk} {\bf J^\dag_k \Phi_k}.
\end{align}

The gains and leakages are generally computed using observations of calibrators, or can be fit to data along with source structure; examining the measured correlation matrix reveals useful structures that illuminate their impact. First we introduce the measured single-station pre-gain fields $R'_D$ and $L'_D$, which have a small contribution from the orthogonal handedness given by the so-called ``$D$ terms'' $D_R$ and $D_L$. These leakage-affected fields also carry the only non-canceling term from the geometric field rotation, manifesting as a rotation of the orthogonal handedness:
\begin{align}
    R'_D &= R+L D_R  e^{i 2 \Phi},\\
    L'_D &= L+R D_L  e^{i 2 \Phi}.
\end{align}
The fully corrupted $\rho'_{jk}$ can then be expressed simply in terms of the gains at each station and the leakage-affected fields:
\begin{align}
    \rho'_{jk} &= 
    \begin{bmatrix}
        G_{R,j} G^*_{R,k} R'_{D,j}R'^*_{D,k} & G_{R_j} G_{L,k}^* R'_{D,j} L'^*_{D,k}\\
        G_{L,j} G_{R,k}^* L'_{D,j} R'^*_{D,k} & G_{L,j} G_{L,k}^* L'_{D,j}L'_{D,k}.
    \end{bmatrix}
\end{align}
The measured correlation products then correspond to the elements of the corrupted correlation matrix: 
\begin{align}
    R'_j R'^*_k &= G_{R,j} G_{R,k}^* R'_{D,j}R'^*_{D,k},\\
    R'_j L'^*_k &= G_{R,j} G_{L,k}^* R'_{D,j} L'^*_{D,k},\\
    L'_j R'^*_k &= G_{L,j} G_{R,k}^* L'_{D,j} R'^*_{D,k},\\
    L'_j L'^*_k &= G_{L,j} G^*_{L,k} L'_{D,j}L'_{D,k}.
\end{align}

As is apparent from \autoref{eq:rho}, measured Stokes visibilities may be constructed by taking linear combinations of elements of $\rho'_{jk}$ as follows:
\begin{align}
    \tilde{I}'_{jk} = \frac{R'_j R'^*_k + L'_j L'^*_k}{2},\\
    \tilde{Q}'_{jk} = \frac{L'_j R'^*_k + R'_j L'^*_k}{2},\\
    \tilde{U}'_{jk} = i \frac{L'_j R'^*_k - R'_j L'^*_k}{2},\\
    \tilde{V}'_{jk} = \frac{R'_j R'^*_k - L'_j L'^*_k}{2}.
\end{align}
We are interested in ratios between linear polarimetric Stokes visibilities and the Stokes $I$ visibility. Both $\breve{m}$ and $\breve{\beta}_2$ are linearly related to quotients $\tilde{Q}/\tilde{I}$ and $\tilde{U}/\tilde{I}$; $\tilde{Q}$ and $\tilde{U}$ are similarly just linear combinations of the same correlation matrix elements, so without loss of generality, we will consider only the ratio $\tilde{Q}/\tilde{I}$. We define the complex gain ratio $g \equiv G_R/G_L$, and find
\begin{align}
    \frac{\tilde{Q}'_{jk}}{\tilde{I}'_{jk}} &= \frac{L'_j R'^*_k + R'_j L'^*_k}{R'_j R'^*_k + L'_j L'^*_k}\nonumber\\
    &= \frac{G_{R,j} G_{L,k}^* R'_{D,j}  L'^*_{D,k} + G_{L,j} G_{R,k}^* L'_{D,j} R'^*_{D,k} }{G_{R,j} G^*_{R,k} R'_{D,j} R'^*_{D,k} + G_{L,j} G^*_{L,k} L'_{D,j} L'^*_{D,k} },\nonumber\\
    &= \frac{G_{R,j} G_{L,k}^*\left(R'_{D,j}  L'^*_{D,k} + g_j^{-1} g_k^* L'_{D,j} R'^*_{D,k} \right)}{G_{L,j} G_{R,k}^*\left(g_j R'_{D,j} R'^*_{D,k} +  g_k^{*-1} L'_{D,j} L'^*_{D,k} \right)},\nonumber\\
    &= \left(\frac{g_j R'_{D,j}  L'^*_{D,k} + g_k^* L'_{D,j} R'^*_{D,k} }{g_j g_k^*  R'_{D,j} R'^*_{D,k} + L'_{D,j} L'^*_{D,k} }\right)
\end{align}
We observe that the ratio $\tilde{Q}'_{jk}/\tilde{I}'_{jk}$ depends only on the leakage-affected correlation products and the complex gain ratios $g_j$ and $g_k$. A similar manipulation is possible for $\tilde{U}'_{jk}/\tilde{I}'_{jk}$. Notably, because the left and right gains contain the same atmospheric contribution to the gain phase, the gain ratio cancels any phase variation imposed by atmospheric image translation, the primary corruption to VLBI phases.

The complex gain ratio can be decomposed into two unknowns, the gain amplitude ratio $|g| =|G_R|/|G_L|$ and the right left gain phase offset $\angle g = \angle G_R - \angle G_L$. Each of these quantities is routinely calibrated much more easily than the absolute values $|G|$ or $\angle G$, as measurements of bright, unresolved, unpolarized and linearly polarized calibrator sources constrain the full complex ratio while leaving both $|G|$ and $\angle G$ unspecified in pessimal cases \citep{PaperIII}. Thus, polarimetric ratios are significantly more robust than, for example, visibility amplitudes, in the limit when the $D$ terms are small. These leakage effects are typically a few percent, reaching $10\%$ in pessimal cases, but can typically be modeled during data analysis as was done in \citetalias{PaperVII}.


\section{Relative Signs of Bessel Functions}
\label{sec:bessel}
\begin{figure}[ht!]
    \centering
    \includegraphics[width=0.49\textwidth]{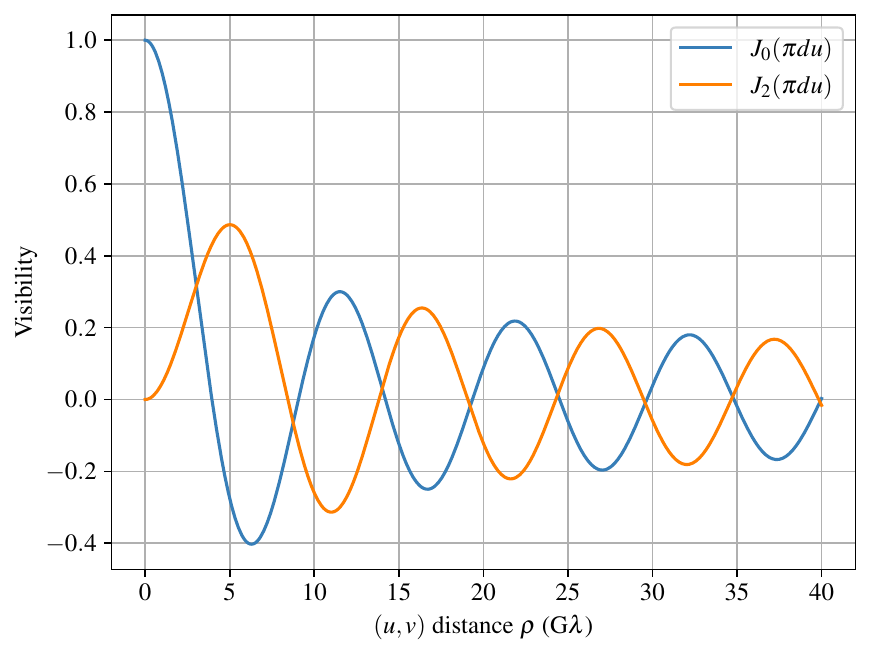}
    \includegraphics[width=0.49\textwidth]{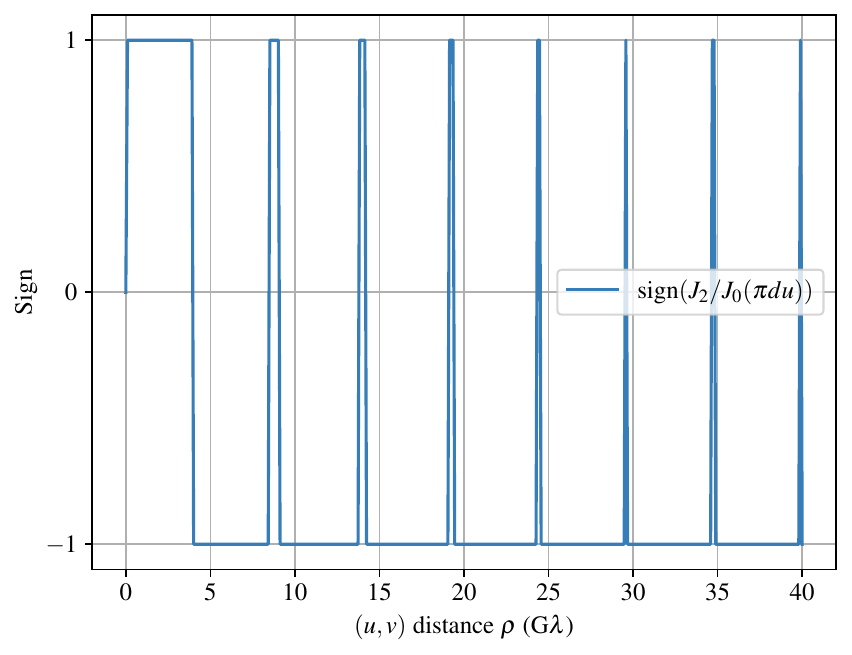}
    \caption{Left: plots of the Bessel functions $J_0$ and $J_2$ corresponding to the interferometric response to rotationally symmetric Stokes $I$ and $Q+iU$ images, respectively, assuming $d=40\mu$as. Right: relative sign of the two functions, which is constant after the first null of $J_0$ except for an asymptotically shrinking region between the nulls of $J_0$ and $J_2$.}
    \label{fig:bessel}
\end{figure}

A useful property of Bessel functions of the first kind is that functions with same-parity order (that is, all even and all odd) have asymptotically close zeros with increasing argument. \citet{McMahon_1894} obtained the following approximate formula for the $s^{\rm th}$ root of the equation $J_n(x)=0$ (here showing the first two terms of McMahon's equation 8 with a slight change in notation):
\begin{align}
    x_n^{(s)} &= Y - \frac{4n^2-1}{8Y} + \ldots,\\
    Y &\equiv \frac{1}{4}\pi(2 n -1 + 4s).
\end{align}
For the thin ring image, we are interested in the separation between the $(s+1)^{\rm th}$ root of $J_0$ and the $s^{\rm th}$ root of $J_2$ (because $J_2$ has a root at $x=0$ and is thus ``one root ahead''):
\begin{align}
    \label{eq:nullsep}
    x_0^{(s+1)} - x_2^{(s)} &\approx \frac{8}{4 \pi s + 3 \pi}.
\end{align}
For large $s$, the null separation falls like $1/s$, clearly approaching zero. For small $s$, the null separation is approximately $8/3\pi < 1$, which is a small fraction of the $\sim \pi$ null spacing of each individual function.

In the particular case of the EHT coverage of \m{}, the second null of $J_0$ is approached by the longest baselines at 230 GHz and will be exceeded by the longest baselines at 345 GHz and beyond; this is sufficient for the nulls of $J_0$ and $J_2$ to nearly align, meaning that the quotient of the two functions will have nearly constant (negative) sign except for brief transitional regions around each null.

As shown in \autoref{fig:bessel}, the relative signs (here standing in for the relative phase between $\tilde{P}$ and $\tilde{I}$) between $J_2$ and $J_0$ are remarkably stable, increasingly so with larger $(u,v)$ distance. Meanwhile, the separation in nulls falls rapidly, following the approximate behavior derived in \autoref{eq:nullsep}.

\section{Example Sites}
\label{sec:sites}

\begin{figure*}[ht!]
    \centering
    \includegraphics[width=\textwidth]{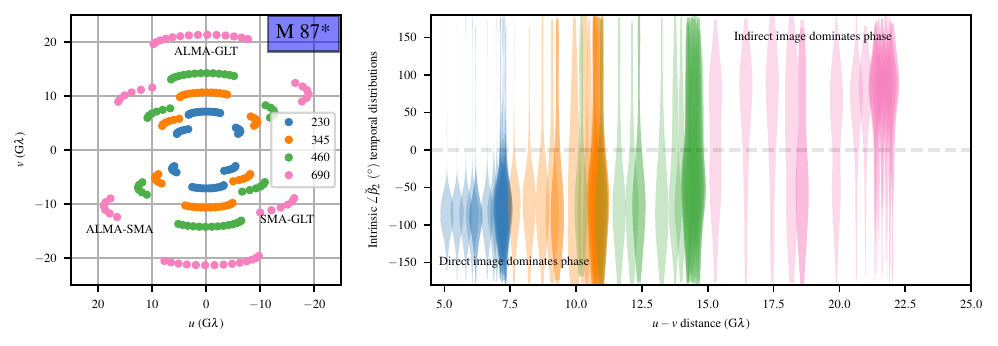}
    \includegraphics[width=\textwidth]{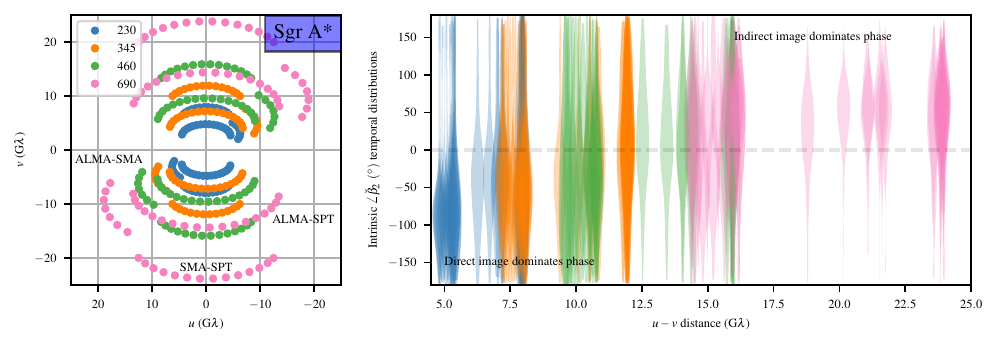}
    \caption{Top left: $(u,v)$ coverage from ALMA, GLT, and SMA at 230, 345, and 690 GHz. Top right: distributions of $\angle \breve{\beta}_2$ at each $(u,v)$ point over the duration of the \m{} simulation. Bottom row: same as top row, but for Sgr A*, using ALMA, SPT, and SMA. Observations are computed assuming a perfect, noiseless instrument observing every twenty minutes. The temporal variation is increased when the visibility response to a $\sim40\mu$as ring passes through nulls, as well as when the $n=0$ and $n=1$ contributions are comparable (around 15 G$\lambda$). However, the longest baselines at 690 GHz still have a mean value that is clearly distinguishable in sign from the short-baseline average. Moreover, the spiral quotient itself permits coherent averaging over long timescales, critical to the success of the experiment.}
    \label{fig:violins}
\end{figure*}

We will examine the Fourier coverage available at 230, 345, 480, and 690 GHz from the Earth's surface at likely participant sites in high-frequency observations. We consider four locations with existing or planned facilities: the Atacama Large (sub)Millimeter Array (ALMA), the Greenland Telescope (GLT, at its planned summit location), the Submillimeter Array (SMA), and the South Pole Telescope (SPT). We do not consider detailed instrument properties in this letter, instead using only geographical position. Thus, the ALMA location is equally indicative of requirements for the Atacama Pathfinder Experiment (APEX) telescope, and the SMA is predictive for the James Clark Maxwell Telescope (JCMT).

These sites are unique in that few other locations on Earth are ever capable of 690 GHz VLBI at the sensitivities relevant to this observation. For these particular sites, we consider two potential limitations on observations: intrinsic source variation and instrument noise.

\autoref{fig:violins} shows the multifrequency coverage provided by the sites under consideration along with the corresponding temporal distributions of $\angle\breve{\beta}_2$ over 1000 gravitational times ($G M / c^3$) in the simulation as ray traced for \m{} and \s{}, where $G$ is the gravitational constant, $M$ is the black hole mass, and $c$ is the speed of light. Instrumental noise is not considered; all variation represents intrinsic evolution of source structure. The phase variation is broadly consistent with the temporal variation seen in the image domain in Figure 2 of \citetalias{Palumbo_2022}, with the exception of regions near nulls of the visibility response and transitions between $n=0$ and $n=1$.

\bibliography{main}

\end{document}